\documentclass[aps,prd,twocolumn,showpacs,10pt,superscriptaddress,preprintnumbers,nofootinbib]{revtex4-2}
\usepackage{amsmath, amssymb}
\usepackage{physics, bm}
\usepackage{graphicx}
\usepackage{xcolor}
\usepackage[colorlinks, linkcolor=blue!80!black, citecolor=green!50!black, urlcolor=magenta]{hyperref}

\usepackage[normalem]{ulem}
\usepackage{soul}
\usepackage{multirow}
\usepackage{orcidlink}

\newcommand{\beq}[1][]{\begin{equation}\label{#1}}
\newcommand{\eeq}{\end{equation}}
\newcommand{\bea}{\begin{eqnarray}}
\newcommand{\eea}{\end{eqnarray}}
\newcommand{\nn}{\nonumber}

\newcommand{\fig}[1]{Fig.~\ref{#1}}

\newcommand{\eq}[1]{Eq.~\eqref{#1}}
\newcommand{\eqs}[2]{Eqs.~\eqref{#1} and~\eqref{#2}}

\newcommand{\refs}[1]{Refs.~\cite{#1}}

\newcommand{\pp}[1]{\left(#1\right)}
\newcommand{\bigpp}[1]{\big(#1\big)}
\newcommand{\bb}[1]{\left[#1\right]}

\newcommand{\vv}[1]{\left\langle #1 \right\rangle}

\newcommand{\M}{\mathcal{M}}

\begin{document}

\preprint{
	{\vbox {
		\hbox{\bf JLAB-THY-24-4236}	
		\hbox{\bf MSUHEP-24-022}
}}}
\vspace*{0.2cm}

\title{Transverse spin effects and light-quark dipole moments at lepton colliders}

\author{Xin-Kai Wen\,\orcidlink{0009-0008-2443-5320}}
\email{xinkaiwen@pku.edu.cn}
\affiliation{School of Physics, Peking University, Beijing 100871, China}
\affiliation{Institute of High Energy Physics, Chinese Academy of Sciences, Beijing 100049, China}
\affiliation{China Center of Advanced Science and Technology, Beijing 100190, China}

\author{Bin Yan\,\orcidlink{0000-0001-7515-6649}}
\email{yanbin@ihep.ac.cn (corresponding author)}
\affiliation{Institute of High Energy Physics, Chinese Academy of Sciences, Beijing 100049, China}

\author{Zhite Yu\,\orcidlink{0000-0003-1503-5364}}
\email{yuzhite@jlab.org (corresponding author)}
\affiliation{Theory Center, Jefferson Lab, Newport News, Virginia 23606, USA}

\author{C.-P. Yuan\,\orcidlink{0000-0003-3988-5048}\,}
\email{yuanch@msu.edu}
\affiliation{Department of Physics and Astronomy, Michigan State University, East Lansing, Michigan 48824, USA}

\begin{abstract}
We propose to probe light-quark dipole interactions at lepton colliders using the azimuthal asymmetry of a collinear dihadron pair $(h_1h_2)$ produced in association with another hadron $h'$. This asymmetry, arising from quantum interference in the quark spin space, is exclusively sensitive to dipole interactions at the leading power of the new physics scale and simultaneously probes both the real and imaginary components of the dipole couplings. By combining all possible channels of $h'$, this method allows for disentangling the up and down quark dipole moments and has the potential to significantly strengthen current constraints by one to two orders of magnitude.
\end{abstract}

\maketitle

\section{Introduction}
Measuring electroweak dipole moments of particles is crucial for testing the Standard Model (SM) and probing New Physics (NP). 
Electroweak dipole moments are related to $CP$ violation, needed to generate the observed baryon asymmetry of the Universe~\cite{Sakharov:1967dj},
and can be systematically parametrized by dimension-six operators within the framework of the SM effective field theory (SMEFT)~\cite{Buchmuller:1985jz,Grzadkowski:2010es}.
These operators introduce NP interactions at the scale $\Lambda$, 
which are characterized by fermion helicity-flip effects in high-energy scattering 
and could be the origin of certain observed anomalies at the Large Hadron Collider (LHC). 
One example is the observed Lam-Tung relation breaking in the Drell-Yan process~\cite{Li:2024iyj, Lam:1978pu, Lam:1978zr, Lam:1980uc, ATLAS:2016rnf, CMS:2015cyj, Gauld:2017tww, Frederix:2020nyw}. 
However, because of this property, in the current global analyses of SMEFT operators that only use unpolarized production rates~\cite{Escribano:1993xr,Alioli:2018ljm, daSilvaAlmeida:2019cbr,Boughezal:2021tih,Cao:2021trr,Shao:2023bga,Bonnefoy:2024gca}, 
to which the contributions from light-fermion dipole operators start only quadratically at $\mathcal{O}(1/\Lambda^4)$ 
or are significantly suppressed by the small fermion mass(es) in their interference with the SM amplitudes at $\mathcal{O}(1/\Lambda^2)$,
the corresponding Wilson coefficients are poorly constrained.

To resolve this issue, there have recently been proposals of new observables to make use of the transverse spin information~\cite{Wen:2023xxc, Boughezal:2023ooo, Wang:2024zns, Wen:2024cfu}.
These observables are particularly effective because the transverse spin of a light fermion is sensitive to the interference 
between the electroweak dipole and the SM interactions, so occur at $\mathcal{O}(1/\Lambda^2)$ with no suppression from the fermion mass.
While this can be done relatively easily for lepton beams by controlling their polarization with the guiding magnetic field of storage rings~\cite{Wen:2023xxc}, 
the transverse spins of light quarks are not directly accessible due to color confinement.
They have to be indirectly controlled through the quark transversity distribution $h_q(x,\mu)$ from initial-state transversely polarized proton beam~\cite{Boughezal:2023ooo, Wang:2024zns},
or observed via the azimuthal asymmetry of a hadron pair in the final-state fragmentation process~\cite{Wen:2024cfu}.
Both methods have been theoretically investigated to constrain the dipole moments of light quarks in electron-proton ($ep$) collisions 
at the upcoming Electron-Ion Collider~\cite{Boughezal:2023ooo,Wen:2024cfu}.
\footnote{The light quark dipole interactions can also be constrained by the nucleon dipole moment measurements, 
but interpreting the results inevitably relies on nonperturbative input of the nucleon spin structure~\cite{Hecht:2001ry, Pitschmann:2014jxa, Liu:2017olr, ParticleDataGroup:2022pth, Sahoo:2016zvr, Bhattacharya:2012bf, Mena:2024qou, Bicudo:1998qb}.}
However, due to the available final states in $ep$ collisions, the constraints are limited only to a specific quark flavor combination. 
Hence, it requires a combined analysis with other observables to disentangle the flavors.

It is the purpose of this paper to examine the possibility of constraining light-quark dipole couplings at a future lepton-lepton collider.
This can be viewed as a crossed process of dihadron production in $ep$ collisions, where the initial-state proton is flipped to the final state, with a  much cleaner background. 
Furthermore, the experimental observable is no longer limited to the proton-dihadron final state.
As we will show, 
a combined analysis of the associated production of a  $(\pi^+\pi^-)$ pair with another hadron $h'$,
e.g., $h' = \pi^{\pm}$, $K^{\pm}$, $p$ or $\bar{p}$,
enables us to disentangle the dipole couplings of up and down quarks.

\section{Observable and factorization}
We consider the inclusive associated production of a dihadron-hadron pair in $e^-e^+$ collisions, in the center-of-mass (c.m.) frame,
\begin{equation}\label{eq:sia}
	e^{-}(\ell) + e^{+}(\ell^\prime)  \rightarrow  [h_1(p_1) + h_2(p_2) ] + h'(p') + X,
\end{equation}
where the three hadrons $h_1$, $h_2$, and $h'$ all have large energies,
with the pair $(h_1, h_2)$ close to each other in angles but widely separated from $h'$.
Physically, this scattering is dominated by the partonic process where $(h_1, h_2)$ arise from the fragmentation of a single parton and $h'$ from another.
Although partons cannot be directly observed, the kinematic distribution of the observed hadrons is dictated by the spin state of the fragmenting parton. 
In particular, by measuring two hadrons in a single jet, 
a nontrivial azimuthal asymmetry with respect to the parton direction can arise from the transverse spin of the parton~\cite{Collins:1993kq, Artru:1995zu, Jaffe:1997hf, Bianconi:1999cd,Barone:2001sp, Bacchetta:2008wb, Metz:2016swz},
which is sensitive to the dipole interaction that the parton can potentially involve~\cite{Wen:2024cfu}.

\begin{figure}[htbp]
	\centering
	\includegraphics[scale=0.4]{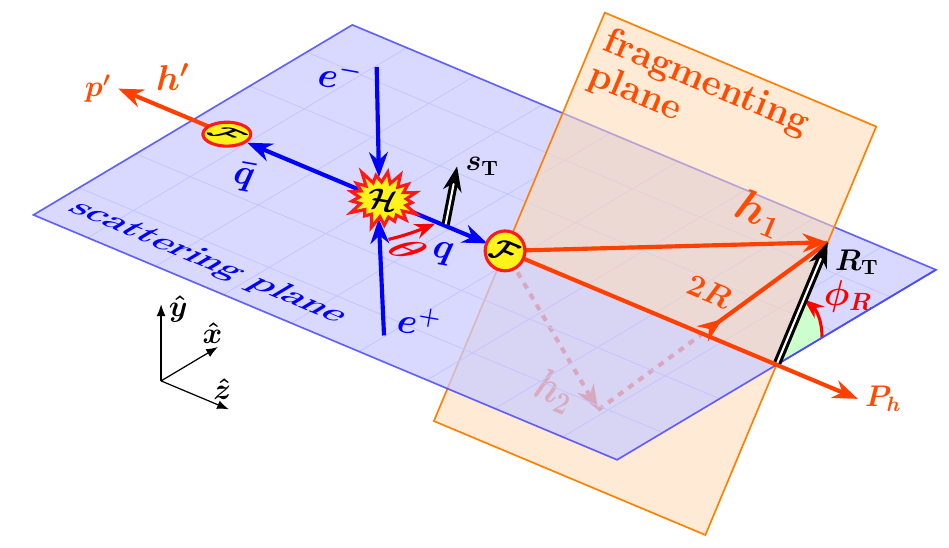}
	\caption{LO kinematic configuration of associated dihadron-hadron production. 
	The quark $q$ and antiquark $\bar{q}$ fragment into the dihadron pair $(h_1h_2)$ and hadron $h'$, respectively.
	}
	\label{Fig:Geo}
\end{figure}

Using $P_h = p_1 + p_2$, we define a local coordinate system
oriented around the dihadron on the production plane,
\beq
	\hat{z} = \frac{\bm{P}_h}{|\bm{P}_h|}, \quad
	\hat{y} = \frac{\bm{\ell} \cross \hat{z}}{|\bm{\ell} \cross \hat{z}|}, \quad
	\hat{x} = \hat{y} \cross \hat{z},
\label{eq:frame}
\eeq
with respect to which we define the transverse component
$\bm{R}_T = (R_T^x, R_T^y)$ of the momentum difference $R^\mu = (p_1^\mu - p_2^\mu) / 2$
and its azimuthal angle $\phi_R$, as shown in Fig.~\ref{Fig:Geo}.
It is the distribution of $\phi_R$ that is of our interest.
We also define $q_{\ell} = \ell + \ell'$ together with the kinematic invariants,
\begin{equation*}
	s = q_{\ell}^2, \quad 
	y = \frac{P_h \cdot \ell }{P_h \cdot q_{\ell}}, \quad 
	z = \frac{2 P_h \cdot q_{\ell}}{s}, \quad 
	\bar{z} = \frac{2 p' \cdot q_{\ell}}{s},
\end{equation*}
where $y = (1 - \cos\theta) / 2$ trades for the polar angle $\theta$ of the dihadron,
and $(z, \bar{z}) = 2 (P_h^0, p^{\prime 0}) / \sqrt{s} $ are the fractional energies. 
Then we can write the cross section of Eq.~\eqref{eq:sia} in a factorized form~\cite{Collins:1993kq,Collins:2011zzd},
\begin{align}
	&\frac{d\sigma}{dy \, dz \, d\bar{z} \, dM_h \, d\phi_R} 
	= \frac{1}{32\pi^2 s} \sum_{q, \, q\to \bar{q}} C_q(y) \, D^{h'}_{\bar{q} }(\bar{z}) \nn\\
	& \times \big[D^{h_1 h_2}_{q}(z, M_h)
	 -(\bm{s}_{T,q}(y)\times \hat{\bm R}_T)^z H^{h_1 h_2}_{q}(z, M_h) \big],
\label{eq:diXsec}
\end{align}
which holds in the kinematic region with $M_h \ll |\bm{P}_h|$, $m_{h'} \ll |\bm{p}'|$, and $M_h$, $m_{h'} \ll \sqrt{(P_h + p')^2}$,
where $M_h = \sqrt{P_h^2}$ and $m_{h'}$ are the invariant masses of the dihadron $(h_1 h_2)$ and hadron $h'$, respectively.

\begin{figure}[htbp]
	\centering
	\includegraphics[scale=0.6]{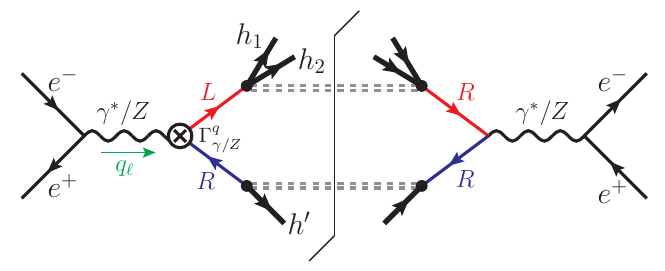}
	\caption{LO cut diagram representation of the factorization in Eq.~\eqref{eq:diXsec}. 
	The gray dashed double lines represent anything else produced from the quark or antiquark fragmentation.
	We have indicated the chirality `$L$' or `$R$' of quark lines. At the cross vertex is the dipole interaction that flips the chirality.
	}
	\label{Fig:MM}
\end{figure}

The factorization formula in Eq.~\eqref{eq:diXsec} separates the physical stages at different scales, as depicted in Fig.~\ref{Fig:MM}.
At the leading order (LO)
\footnote{Higher-order QCD corrections are not expected to alter the conclusions of our analysis, 
analogous to the cancellation effect observed in the spin asymmetry ratio for the polarized Drell-Yan process~\cite{deFlorian:2017ogw}.  
While dihadron pairs can also arise from polarized gluons, 
such contributions can only generate $\cos2\phi_R$ and/or $\sin2\phi_R$ modulations. 
Consequently, the $\cos\phi_R$ and $\sin\phi_R$ asymmetries represent 
robust signatures of transversely polarized quarks---an attribute unaffected by gluon radiation due to chiral symmetry.}
the hard scattering produces a quark and antiquark, $e^-_{\lambda_1} e^+_{\lambda_2} \to q_{\lambda} \bar{q}_{\bar{\lambda}}$, 
whose helicity amplitude $\M^{\lambda_1 \lambda_2}_{\lambda \bar{\lambda}}$ determines the hard coefficient $C_q$, 
which controls the unpolarized production rate, 
and the transverse spin $\bm{s}_{T,q} = (s^x_q, s^y_q)$ of the quark through the (unnormalized) density matrix, 
\beq
	W^q_{\lambda\lambda'} \equiv \frac{1}{4} \M^{\lambda_1 \lambda_2}_{\lambda \bar{\lambda}} \bigpp{ \M^{\lambda_1 \lambda_2}_{\lambda' \bar{\lambda}} }^*
		= \frac{C_q}{2} \pp{ \delta_{\lambda\lambda'} + s_q^i \sigma^i_{\lambda\lambda'} },
\label{eq:Wq}
\eeq
where the $\lambda$'s are the helicity indices with repeated ones summed over, 
$\sigma^i = (\sigma^x, \sigma^y, \sigma^z)$ are the Pauli matrices,
and $s_q^i = (s_q^x, s_q^y, s_q^z)$ is the quark spin vector defined in the dihadron frame, cf.~Eq.~\eqref{eq:frame}.
In the case when the initial-state leptons are polarized, we shall average their helicities in Eq.~\eqref{eq:Wq} with proper density matrices.

Then, the polarized quark fragments into the dihadron pair, 
with its transverse spin causing an azimuthal asymmetry in the pair, as given by
\beq
	( \bm{s}_{T, q} \times \hat{\bm{R}}_T )^z = s^x_{q} \sin\phi_R - s^y_{q} \cos\phi_R
\label{eq:spin}
\eeq
in the dihadron frame, where $\hat{\bm{R}}_T = \bm{R}_T / |\bm{R}_T| = (\cos\phi_R$, $\sin\phi_R)$.
With the dihadron system described by their total longitudinal momentum fraction $\xi$ in the quark and invariant mass $M_h$,
the fragmentation is quantified by two nonperturbative dihadron fragmentation functions (DiFFs), 
$D^{h_1 h_2}_{q}(\xi, M_h)$ and $H^{h_1 h_2}_{q}(\xi, M_h)$,
where $D^{h_1 h_2}_{q}$ controls the rate and the ratio $H^{h_1 h_2}_{q} / D^{h_1 h_2}_{q}$ plays the role of quark spin analyzing power. 
Similarly, the fragmentation of the antiquark into a single hadron is described 
by an unpolarized fragmentation function (FF), $D^{h'}_{\bar{q} }(\zeta)$, 
with $\zeta$ being the longitudinal momentum fraction of $h'$ in $\bar{q}$.
In Eq.~\eqref{eq:diXsec}, we have used the LO kinematics to set $\xi = z$ and $\zeta = \bar{z}$,
and have chosen the factorization scale $\mu$ to be at $\sqrt{s}$, whose dependence has been suppressed for simplicity.

\section{Dipole interaction and transverse spin}
As shown in Eq.~\eqref{eq:Wq}, the transverse spin $\bm{s}_{T,q}$ of our interest is given by the off-diagonal element of the density matrix,
\beq[eq:W-+]
	W^q_{-+} = \frac{1}{4} \bb{ \M^{\lambda_1 \lambda_2}_{-+} \bigpp{ \M^{\lambda_1 \lambda_2}_{++} }^*
		+ \M^{\lambda_1 \lambda_2}_{--} \bigpp{ \M^{\lambda_1 \lambda_2}_{+-} }^* },
\eeq
which is the interference of amplitudes that differ by a single quark helicity flip.
In the SM with the light quark masses neglected, such interference is forbidden due to the chiral symmetry---every SM vertex conserves the quark chirality, and thus, helicity.
This fact, on the other hand, makes $\bm{s}_{T,q}$ a very sensitive probe to NP dipole interactions which flip the quark chirality.

Within the SMEFT framework~\cite{Buchmuller:1985jz,Grzadkowski:2010es},
we parametrize the electroweak light quark dipole interactions as
\begin{align}
	&\mathcal{L}_{\rm eff} = -\frac{1}{\sqrt{2}} \bar{q}_L \sigma^{\mu\nu}
	\left(g_1 \Gamma_B^d B_{\mu\nu}+g_2 \Gamma_W^d \sigma^a W_{\mu\nu}^a \right) \frac{H}{v^2} d_R
	\nn\\
	&\hspace{1em} -\frac{1}{\sqrt{2}} \bar{q}_L \sigma^{\mu\nu}
	\left(g_1 \Gamma_B^u B_{\mu\nu}+g_2 \Gamma_W^u \sigma^a W_{\mu\nu}^a \right) \frac{\widetilde{H}}{v^2} u_R
	+{\rm H.c.},
\label{eq:Leff}
\end{align}
where $q_L = (u_L, d_L)^T$ is the first-generation left-handed quark doublet, 
$u_R$ and $d_R$ are the right-handed singlets,
and $H$ and $\widetilde{H} = i \sigma^2 H^*$ are the Higgs doublets, with $v=246~{\rm GeV}$ being the vacuum expectation value. 
The $B_{\mu\nu}$ and $W_{\mu\nu}^a$ are the field strength tensors of the gauge groups U$(1)_Y$ and SU$(2)_L$, respectively, 
and $g_1$ and $g_2$ are the corresponding gauge couplings. 
The dimensionless Wilson coefficients $\Gamma_B^{u,d}$ and $\Gamma_W^{u,d}$ are the dipole interaction couplings. 
For convenience, we redefine the $u$ and $d$ quark dipole couplings in terms of the gauge boson mass eigenstates as 
$\Gamma_\gamma^{u, d} = \mp \Gamma_W^{u, d} - \Gamma_B^{u, d}$ and $\Gamma_Z^{u, d} = \mp c_W^2\Gamma_W^{u, d} + s_W^2\Gamma_B^{u, d}$,
where the ``$\mp$'' corresponds to $u$ or $d$, respectively, and  $s_W\equiv\sin\theta_W$ and $c_W\equiv\cos\theta_W$, with $\theta_W$ being the weak mixing angle. 

The Lagrangian in \eq{eq:Leff} conserves the charge conjugation symmetry, 
but breaks the parity and $CP$ symmetries due to the phases of the dipole couplings $\Gamma^q_{\gamma, Z}$.
Since $\Gamma^q_{\gamma, Z}$ and $\Gamma^{q*}_{\gamma, Z}$ flip the chirality in opposite ways,
one can easily read from \eqs{eq:W-+}{eq:Leff} that $W^q_{-+} \propto \Gamma^q_{\gamma, Z}$, 
which originates from the interference of the dipole and the SM interactions, as illustrated in \fig{Fig:MM}.
So we write
\beq
	W^q_{-+}(y) = \Gamma^q_{\gamma} \, w^q_{\gamma}(y) + \Gamma^q_{Z} \, w^q_{Z}(y),
\eeq
with a similar relation for the antiquark.
At the LO, the two coefficients $w^q_{\gamma, Z}$ are real functions of the hard scattering kinematics,
\footnote{We neglect the imaginary part of $w_{\gamma,Z}^q$ from the $Z$-width, as it is suppressed.}
so the transverse spins are simply given by
\begin{align}
	C_q \, s^x_q &= 2 \Re W^q_{-+}
		= 2 \left( w^q_{\gamma} \Re\Gamma^q_{\gamma} + w^q_Z \Re\Gamma^q_Z \right), \nn\\
	C_q \, s^y_q &= 2 \Im W^q_{-+}
		= 2 \left( w^q_{\gamma} \Im\Gamma^q_{\gamma} + w^q_Z \Im\Gamma^q_Z \right),
\label{eq:w-G}
\end{align}
which involve real and imaginary parts of the dipole couplings, respectively.

By measuring the quark transverse spin through the $\phi_R$ modulation in \eq{eq:spin}, 
one gains sensitivity to the dipole couplings---both the real and imaginary parts---at $\mathcal{O}(1 / \Lambda^2)$.
This greatly enhances the sensitivity with only unpolarized rates,
for which one integrates over $\phi_R$ in \eq{eq:diXsec} and only accesses the $C_q \propto |\M^{\lambda_1 \lambda_2}_{\lambda \bar{\lambda}}|^2$,
where the helicity-flipping dipole interactions have to appear in pairs so contribute only quadratically at $\mathcal{O}(1 / \Lambda^4)$.

Now, by performing a parity transformation on the kinematic configuration in \fig{Fig:Geo}, followed by a rotation in the scattering plane by $\pi$~\cite{Yu:2023shd,Wen:2024cfu},
we identify that $s_q^x$ is parity-odd while $s_q^y$ is parity-even, so are their associated $\sin\phi_R$ and $\cos\phi_R$ modulations, respectively.
This means that the spin coefficients $w^q_{\gamma, Z}$ in \eq{eq:w-G} must be parity-odd.
To have nonzero values, they need to depend either on the longitudinal polarization $\lambda_{\ell}$ of the lepton beam 
or on the axial coupling of the $Z$ interaction.
 
The cut diagrams in \fig{Fig:MM} include three types: 
the photon channel squared $(\gamma\gamma)$, the $Z$ channel squared $(ZZ)$, and their interference $(\gamma Z)$.
Their relative weights vary with the c.m.~energy $\sqrt{s}$. 
The $w^q_{\gamma}$ receives contributions from the $\gamma\gamma$ and $\gamma Z$ channels,
so is more suitably probed at low $\sqrt{s}$, where the dominant $\gamma\gamma$ channel requires $\lambda_{\ell} \neq 0$ to give nonzero contribution.
The $w^q_Z$ receives contributions from the $\gamma Z$ and $ZZ$ channels, so is more important at high $\sqrt{s}$, especially near the $Z$-pole, 
where the dominant $ZZ$ channel involves parity-odd couplings $g_A^{e, q}$ and does not require a polarized lepton beam.
 
\section{Azimuthal asymmetry and flavor constraint}
There is currently only the $(h_1 h_2) = (\pi^+\pi^-)$ DiFFs at our disposal, taken from the global-fit results in \refs{Pitonyak:2023gjx, Cocuzza:2023oam, Cocuzza:2023vqs}.
These satisfy the following flavor relations due to isospin and charge conjugation symmetries:
\begin{gather}
	D^{\pi^+\pi^-}_{u} = D^{\pi^+\pi^-}_{d}, \;\;
	H^{\pi^+\pi^-}_{u} = -H^{\pi^+\pi^-}_{d},	\;\;
	H^{\pi^+\pi^-}_{s, \bar{s}, c, \bar{c}, b, \bar{b}} = 0, \nn\\
	D^{\pi^+\pi^-}_{q} = D^{\pi^+\pi^-}_{\bar{q}}, \;\;
	H^{\pi^+\pi^-}_{q} =- H^{\pi^+\pi^-}_{\bar{q}}.
\label{eq:diff-isospin}
\end{gather}
This means that only the $u$ and $d$ quarks can contribute to the azimuthal modulations in \eq{eq:diXsec}.

For simplicity, we integrate over $y$ in \eq{eq:diXsec} and define
\beq[eq:y-int]
	\vv{ C_q } = \int_0^1 dy \, C_q(y), \quad
	\vv{ S_q^i } = \int_0^1 dy \, C_q(y) s_{q}^i(y),
\eeq
with $i = x$ or $y$.
Since the differences between $(C_q, s_q^i)$ and $(C_{\bar{q}}, s_{\bar{q}}^i)$, as derived from \eq{eq:w-G},
are due to the axial coupling $g_A^{q}$ of the $Z$-$q$-${\bar q}$ interaction, 
which breaks charge symmetry and induces a forward-backward asymmetry,
they vanish under the integration of $y$, and hence
$\vv{ C_q } = \vv{ C_{\bar{q}} }$
and 
$\vv{ S_q^i } = \vv{ S_{\bar{q}}^i }$.
This, together with \eq{eq:diff-isospin}, allows us to write \eq{eq:diXsec} in a concise form,
\begin{align}
	&\frac{d\sigma}{dz \, d\bar{z} \, dM_h \, d\phi_R} 
	= \frac{B^0 - B^x \sin\phi_R + B^y \cos\phi_R}{32\pi^2 s},
\label{eq:diXsec-y-int}
\end{align}
where $B^0$ receives contributions only from the SM,
\beq
	B^0 = \sum_{q} \vv{ C_q } D^{\pi^+\pi^-}_q \bigpp{ D^{h'}_q + D^{h'}_{\bar{q}} },
\eeq
whereas $B^{x, y}$ depends linearly on the dipole couplings $\Gamma^q_{\gamma, Z}$ through the transverse spin $\vv{ S^i_{q} }$,
\begin{align}
	B^{i}
	= H^{\pi^+\pi^-}_u 
	\bb{ \vv{ S^i_{u} } \bigpp{ D^{h'}_{\bar{u}} - D^{h'}_u }
		- \vv{ S^i_{d} } \bigpp{ D^{h'}_{\bar{d}} - D^{h'}_d }
	}.
\label{eq:flavor}
\end{align}
It is important to note that $\vv{ S_u^i }$ and $\vv{ S_d^i }$ in \eq{eq:flavor} are multiplied by different coefficients 
that depend on the fragmentation function of the hadron $h'$.
Measurements of different channels of $h'$ associated production provide different constraints on $\Gamma^u_{\gamma, Z}$ and $\Gamma^d_{\gamma, Z}$, which enable them to be separately constrained.

Specifically, we consider the production channels of $h' = \pi^{\pm}$, $K^{\pm}$, or $p / \bar{p}$.
For $h' = \pi^{\pm}$, it is the $(u, \bar{d})$ or $(d, \bar{u})$ flavors whose FFs are dominant, respectively, giving a combination
$\pp{ \vv{ S^i_{u} } + \vv{ S^i_{d} } }$ in $B^i$.
For $h' = K^{\pm}$, with the approximation that $D^{K^{\pm}}_d = D^{K^{\pm}}_{\bar{d}}$, only the coupling $\Gamma^u_{\gamma, Z}$ is probed through $\vv{S^i_u}$.
For $h' = p / \bar{p}$, it is the $(u, d)$ or $(\bar{u}, \bar{d})$ flavors that contribute the most, respectively, leading to a combination 
$\pp{ \vv{ S^i_{u} } - \vv{ S^i_{d} } / 2 }$ in $B^i$, with approximate isospin relation used for the FFs~\cite{Sato:2016wqj,Bertone:2017tyb,Sato:2019yez,Moffat:2021dji,Cocuzza:2022jye,Gao:2024nkz,Gao:2024dbv}, after neglecting the sea quark FF contributions.

That leveraging the associatively produced hadron $h'$ in a multichannel analysis 
enables one to disentangle the quark flavors of the dipole couplings is the most important observation in this paper.
Without measuring $h'$, which amounts to removing the $\bar{z}$ dependence and setting the FF to unity in \eq{eq:diXsec},
the $B^i$ vanishes identically in \eq{eq:flavor}.
To restore the transverse spin effects in this case requires not to integrate over $y$ in \eq{eq:y-int},
and to utilize the forward-backward asymmetry associated with the charge symmetry breaking effects.
However, such observable still constrains only a linear combination of the up and down quark dipole couplings, similarly to \refs{Boughezal:2023ooo,Wen:2024cfu}, but not the individual ones. 

\section{Numerical results and discussion}
To simplify the numerical analysis, we further integrate over $(z, \bar{z}, M_h)$ in Eq.~\eqref{eq:diXsec-y-int},
resulting in a single-differential $\phi_R$ distribution,
\begin{align}
	\frac{2\pi}{\sigma^{h'}} \frac{d\sigma^{h'}}{d\phi_R}
	= 1 - a_R^{h'} \sin\phi_R + a_I^{h'} \cos\phi_R + \order{\Lambda^{-4}},
\label{eq:dis}
\end{align}
which has assumed the $(\pi^+\pi^-)$ dihadron production, with the superscript $h'$ labeling the associated hadron channel.
The normalization factor $\sigma^{h'}$ is the total cross section in the chosen kinematic region.
The azimuthal coefficients $(a_R^{h'}, a_I^{h'})$ are given by the normalized integrations of $(B^x, B^y)$ in \eq{eq:flavor}, 
which are DiFF-weighted average of $(s_q^x, s_q^y)$.
By \eq{eq:w-G}, they are linearly dependent, with the same coefficients, on the real and imaginary parts of the dipole couplings, respectively.
Therefore, measuring $a_R^{h'}$ and $a_I^{h'}$ provides simultaneous and equal constraints on both the real and imaginary parts of the dipole couplings.

It is convenient to introduce two azimuthal asymmetries for measuring $a_R^{h'}$ and $a_I^{h'}$,
\begin{align}
	A_{UD}^{h'}
		&\,= \frac{\sigma^{h'}(\sin\phi_R>0) - \sigma^{h'}(\sin\phi_R<0)}{\sigma^{h'}(\sin\phi_R>0) + \sigma^{h'}(\sin\phi_R<0)}
			= - \frac{2}{\pi} \, a_R^{h'},\nn\\
	A_{LR}^{h'}
		&\,= \frac{\sigma^{h'}(\cos\phi_R>0) - \sigma^{h'}(\cos\phi_R<0)}{\sigma^{h'}(\cos\phi_R>0) + \sigma^{h'}(\cos\phi_R<0)}
			= \frac{2}{\pi} \, a_I^{h'},
\label{eq:SSA}
\end{align}
which can be referred to as ``up-down" and ``left-right" asymmetries, respectively, with respect to the scattering plane in Fig.~\ref{Fig:Geo}.
Here the $\sigma^{h'}(\sin\phi_R \gtrless 0)$ is the integrated cross section with $\sin\phi_R \gtrless 0$, etc.
From the ratio definition in \eq{eq:SSA}, the systematic uncertainties cancel out in these asymmetries, so they will be neglected.
The uncertainties of the asymmetries are then of statistical origin,
\begin{equation}
	\delta A_{LR, UD}^{h'}
	=\sqrt{\frac{1 - (A_{LR, UD}^{h'})^2}{N_{\rm events}}}
	\simeq \frac{1}{\sqrt{N_{\rm events}}},
\label{eq:stats}
\end{equation}
given by the event number $N_{\rm events}$ after applying kinematic cuts at a given collider energy and integrated luminosity. 
Assuming the measurement aligns with the SM, we have made the approximation $A_{LR, UD}^{h'} \simeq 0$ in the second step of \eq{eq:stats}.

The global fits of DiFFs and single-hadron FFs are already available from several groups~\cite{Courtoy:2012ry, Radici:2015mwa, Sato:2019yez, Moffat:2021dji, Cocuzza:2022jye, Sato:2016wqj, Pitonyak:2023gjx, Cocuzza:2023oam, Cocuzza:2023vqs, Bertone:2017tyb, Gao:2024nkz, Gao:2024dbv}.
In our numerical analysis, we utilize the DiFFs and FFs $D^{h'}_q$ from the JAM fits~\cite{Moffat:2021dji, Cocuzza:2022jye, Pitonyak:2023gjx, Cocuzza:2023oam, Cocuzza:2023vqs} 
for $h' = \pi^\pm$ and $K^\pm$, while employing $D^{p/\bar{p}}_q$ from the NNPDF group~\cite{Bertone:2017tyb} as it is not covered by JAM.

As discussed, in order to separate the photon and $Z$ dipole couplings, 
we propose to carry out the measurements both at low-energy lepton colliders, 
such as the Super Tau-Charm Facility (STCF)~\cite{Achasov:2023gey}, BELLE~\cite{Aushev:2010bq},
and high-energy $Z$ factories,
such as the Circular Electron-Position Collider (CEPC)~\cite{CEPCStudyGroup:2018ghi,CEPCStudyGroup:2023quu}, 
International Linear Collider (ILC)~\cite{ILC:2013jhg}, 
and the Future Circular Collider (FCC-ee)~\cite{FCC:2018evy}.
As a benchmark, we choose 
$\sqrt{s} = 10\, {\rm GeV}$ with a polarized electron beam,  with $\lambda_{\ell}=0.8$, to enhance the signal from the $\gamma\gamma$ channel to probe $\Gamma^{u, d}_{\gamma}$,
and $\sqrt{s}=91\, {\rm GeV}$, at the $Z$-pole, with unpolarized lepton beams to probe the $\Gamma^{u, d}_{Z}$ from the $ZZ$ channel,
both with an integrated luminosity of $\mathcal{L}=1000~{\rm fb}^{-1}$.
We select for both cases the optimal kinematic regions, $z \in (0.4, 0.8)$, $\bar{z} \in (0.2, 0.8)$, and $M_h \in (0.4, 1.4)~{\rm GeV}$,
which minimize the uncertainties of the DiFFs in the global-fit analyses~\cite{Bertone:2017tyb, Cocuzza:2022jye, Pitonyak:2023gjx, Cocuzza:2023oam, Cocuzza:2023vqs}
and satisfy the factorization condition in \eq{eq:diXsec}.
Within these regions, we estimate that the relative uncertainties from DiFFs in the spin asymmetries are at the level of  less than 20\%,
with statistical errors being the dominant contribution. Consequently, the DiFF-related uncertainties have negligible impact on our conclusions.
These uncertainties will be further reduced with inclusion of future data from the EIC and lepton colliders~\cite{Zhou:2024cyk}.

\begin{figure}
	\centering
	\includegraphics[scale=0.33]{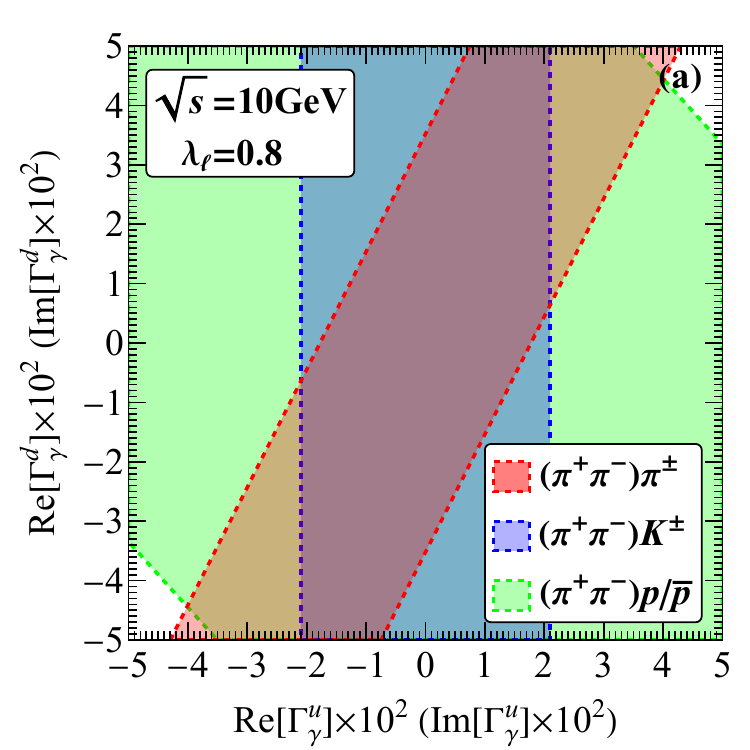}
	\includegraphics[scale=0.33]{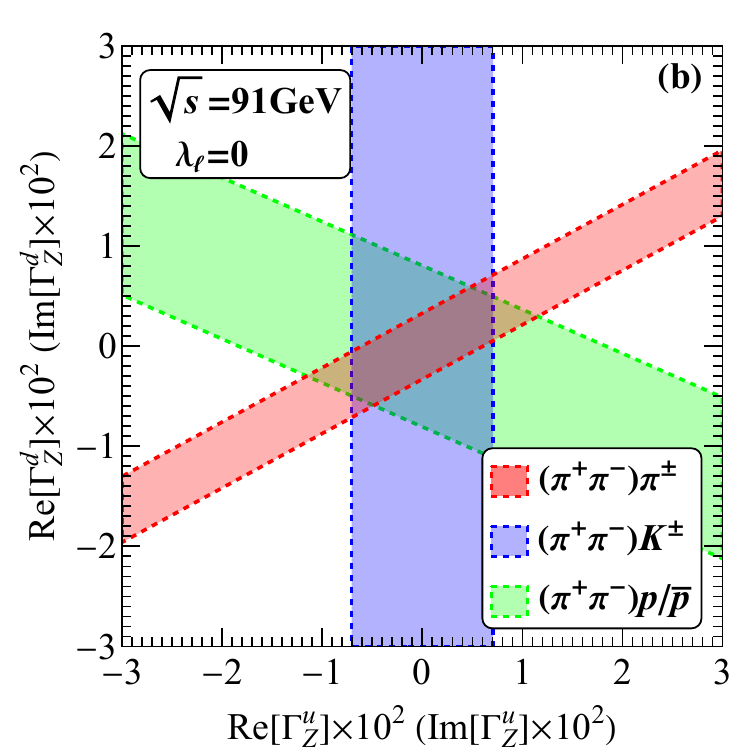}
	\caption{Expected constraints on the real and imaginary parts of the light-quark dipole couplings 
	(a) $\Gamma_{\gamma}^{u,d}$ with $\lambda_\ell=0.8$ at $\sqrt{s}=10~{\rm GeV}$, 
	and (b) $\Gamma_{Z}^{u,d}$ with $\lambda_\ell=0$ at $\sqrt{s}=91~{\rm GeV}$, 
	derived from the azimuthal asymmetries $A_{LR, UD}^{h'}$ for $(\pi^+\pi^-)\pi^\pm$ (red bands), $(\pi^+\pi^-)K^\pm$ (blue bands), and $(\pi^+\pi^-)p/\bar{p}$ (green bands) 
	production processes.
	}
\label{Fig:Limit}
\end{figure}

The projected constraints on the light-quark dipole couplings $(\Gamma_{i}^{u}, \Gamma_{i}^{d})$ are estimated from the asymmetries in Eq.~\eqref{eq:SSA} through a $\chi^2$ analysis, 
where the uncertainties are dominated by the statistical errors from Eq.~\eqref{eq:stats}.
The resulting 68\% confidence level limits are shown in Fig.~\ref{Fig:Limit} for (a) $i = \gamma$ and (b) $i = Z$, 
separately for the $\pi^\pm$ (red bands), $K^\pm$ (blue bands), and $p/\bar{p}$ (green bands) associated production channels.
The real and imaginary parts are constrained in the same way.
As expected, the linear dependence of $A_{LR, UD}^{h'}$ on the dipole couplings leads to straight-band confidence regions;
the $K^\pm$ channel only probes the $\Gamma_{\gamma,Z}^{u}$;
and the $\pi^\pm$ and $p/\bar{p}$ give opposite signs for the slopes of the constraint bands.
These multiple channels significantly constrain the dipole couplings, 
yielding a typical upper limit of $\mathcal{O}(10^{-2})$ for $\Gamma_{\gamma}^{u,d}$ 
and $\mathcal{O}(10^{-3})$ for $\Gamma_{Z}^{u,d}$, simultaneously for their real and imaginary parts.
This complements the constraints from an $ep$ collider, especially 
when comparing with their flat directions of the constraining contour in the $(\Gamma_{i}^{u}, \Gamma_{i}^{d})$ space~\cite{Boughezal:2023ooo,Wen:2024cfu}.

While the dipole couplings $\Gamma_{\gamma, Z}^{u,d}$ can also be constrained by the 
Drell-Yan process, Higgs boson production via vector boson fusion, diboson production, associated production of weak and Higgs bosons at the LHC, 
low-energy data, and electroweak precision observables,
the dipole couplings are only constrained quadratically at $\mathcal{O}(1/\Lambda^4)$ when the transverse spin information associated with the initial or final state of the hard scattering processes is not considered. 
As a result, the sensitivity of those methods, among which Drell-Yan data provide the strongest constraint, is weaker than our results by one to two orders of magnitude, 
even with only one operator considered at a time~\cite{Escribano:1993xr,Alioli:2018ljm, daSilvaAlmeida:2019cbr,Boughezal:2021tih}. 
Furthermore, these methods lack the sensitivity to distinguish between the real and imaginary parts of dipole couplings and cannot separate them from other SMEFT operators. 
This highlights the advantage of the transverse spin observable proposed in this paper, which depends linearly and exclusively on the SMEFT dipole operators.

\section{Conclusion}
In this paper, we introduced a novel approach to probe the light-quark dipole interactions through the azimuthal asymmetry between a collinear pair of hadrons produced in association with a well-separated third hadron at $e^-e^+$ colliders.
This asymmetry results from the transverse spin of a final-state light quark, and is exclusively sensitive to 
the interference of the dipole and the SM interactions at the leading power of $\mathcal{O}(1/\Lambda^2)$, 
unaffected by other new physics interactions in the hard scattering.
This approach can simultaneously constrain the real and imaginary parts of the dipole couplings, provide a new way to probe $CP$ violation effects at $e^-e^+$ colliders, and outperform conventional methods proposed at the LHC and other observables at $e^-e^+$ colliders, by one to two orders of magnitude. 
By leveraging a hadron $h'$ that is produced in association with the dihadron pair and combining all available channels of $h'$ ($\pi^\pm$, $K^\pm$, and $p/\bar{p}$),
we demonstrated---for the first time---that an $e^-e^+$ collider can resolve the flavor degeneracy present at $ep$ colliders, 
enabling separate constraints on the dipole couplings of up and down quarks.
These developments open new avenues for probing NP at the precision frontier and establish an interdisciplinary connection between hadron structure studies and electroweak precision physics.

\section{Acknowledgments}
We thank C.~Cocuzza, A.~Prokudin, and N.~Sato for sharing the dihadron fragmentation function code and helpful discussions.
Xin-Kai Wen is supported in part by the National Natural Science Foundation of China under Grants No.~12235001 and No.~12342502. 
Bin Yan is supported in part by the National Natural Science Foundation of China under Grant No.~12422506, the IHEP under Grant No.~E25153U1 and CAS under Grant No.~E429A6M1.
Zhite Yu is supported in part by the U.S. Department of Energy (DOE) Contract No.~DE-AC05-06OR23177, 
under which Jefferson Science Associates, LLC operates Jefferson Lab,
and C.-P.~Yuan is supported by the U.S. National Science Foundation under Grant No.~PHY-2310291.

\bibliographystyle{apsrev}
\bibliography{reference}

\end{document}